# Providing a way to create balance between reliability and delays in SDN networks by using the appropriate placement of controllers

Amir Javadpour[1,*]
[1]School of Computer Science and Technology, Guangzhou University, Guangzhou, China, 510006
*Correspondence to: a_javadpour@gzhu.edu.cn

**Abstract**
Computer networks covered the entire world and a serious and new development has not formed for many years. But companies and consumer organizations complain about the failure to add new features to their networks and according to their need, like much of the works to be done automatically and they also like to develop and expand their networks on the software side so they don't need new expensive hardware for many of the activities and needs of their network. Analysis of the control layers and writing data in SDN facilitate network management and accelerate innovation in network. In order to develop broad networks of SDN, often a large number controller is needed and located position of controllers in the SDN networks and can be raised as an important and basic issue in the field of SDN network and have impact on reliability of SDN networks. This paper focused on latency and reliability in SDN networks. The latency here means the delay in response to the request of data path that has a significant impact on network latency. In this paper it is shown that the number of controllers and their position can be effective on two measures; reliability and latency in SDN networks.

## 1. Introduction

SDN or software-based networks are trying to increase the intelligence of networks and with the transfer of the control data from the switch and hardware router to virtual software layers of the network and utilizing a centralized software controller provide features such as scheduling[1], scalability[2], flexibility[3], automation[4], intelligence[5] and the software development of network by organizations[6], [7]. SDN is known as the biggest change of computer networks in four decades[8]. SDN first was raised in 2005 and accelerated from 2010. Forming the ONF foundation in 2011 and joining more than eighty largest companies of the networking industry and developing the Open Flow standard was introduced into a new phase[9]. The first products of SDN entered into the market in 2012 and more in 2013 and it is anticipated by 2015, this type of networks are gradually are replacing the traditional networks that are based on Ethernet and TCP / IP[10]. As shown in figure 1, number of forms must be written in the text respectively. As an example, architecture related to software-based networks SDN is shown.

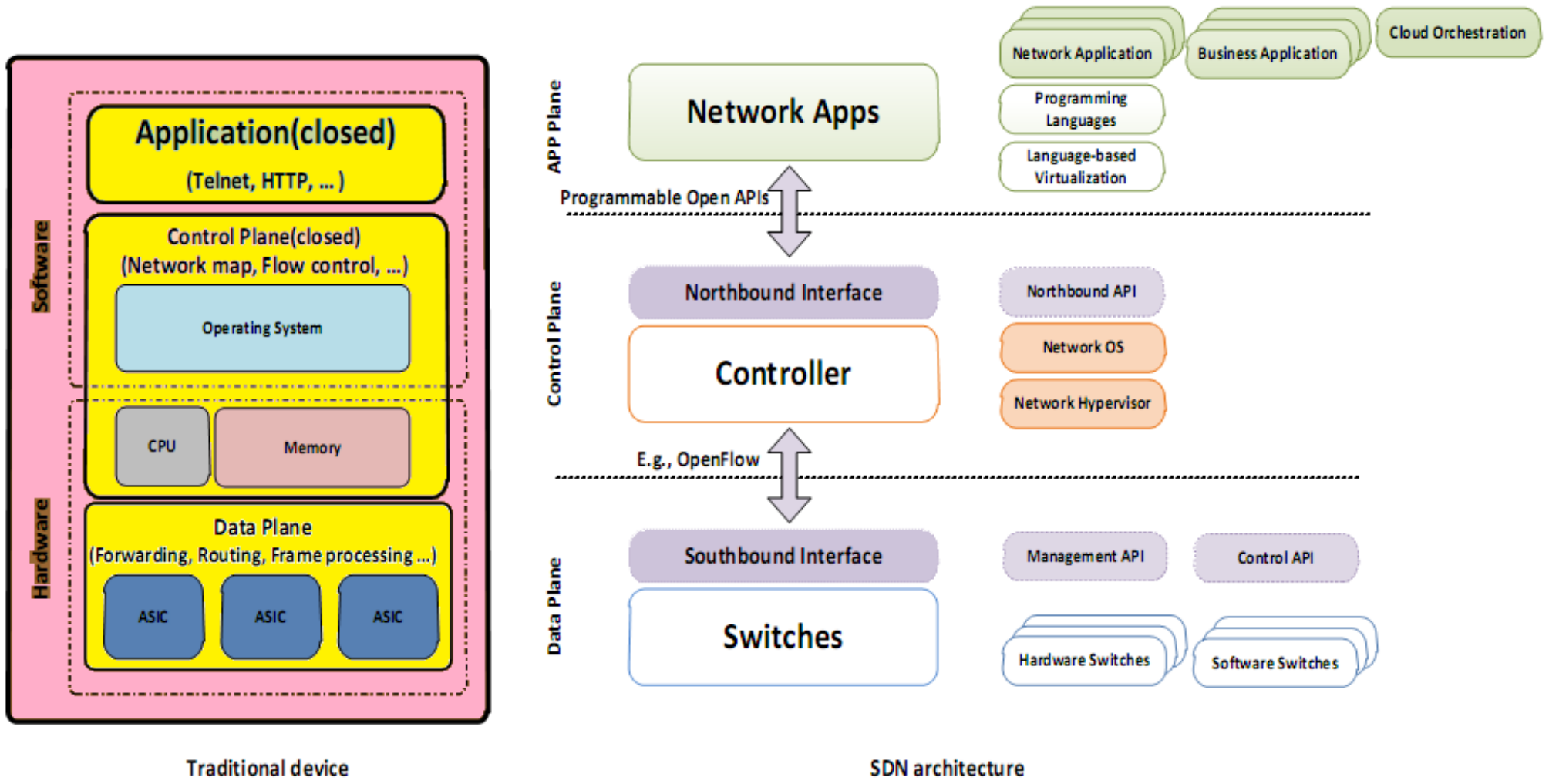

**Fig.1 : Architecture related to software-based networks SDN[9].**

The SDN framework consists of three layers. Infrastructure layer that is called Data plane or write layer that is responsible for data submission and monitoring of local information and statistic collecting of them. Layer above sending or posting data layer is Control plane layer that is responsible for programming and managing posting plans and network function and routing are defined in this layer[11]. This layer contains one or more software controllers that communicate with the elements of the transmission of the network through standard interfaces. This interface is referred to as the interface of southbound levels[12]. Open Flow is one of the interfaces of this field. Open Flow is the first standard communication interface seen in SDN architecture between control and post [13], [14]. The highest layer in SDN framework is called the Application Layer and includes network applications that express the new network properties. This feature includes security and management and schema sending or help control layer in network settings. The application layer receives a global and abstract network view through the controllers and uses that information to provide appropriate guidance to the control layer. The interface between the control layer and the application layer is referred to as the interface of the northbound. The ONF Foundation launched in 2011 with the goal of promoting a new form of SDN networks that is compatible with the Open Flow protocol. Our domain of research in the second layer (figure 2) of the SDN framework is the control layer[9].

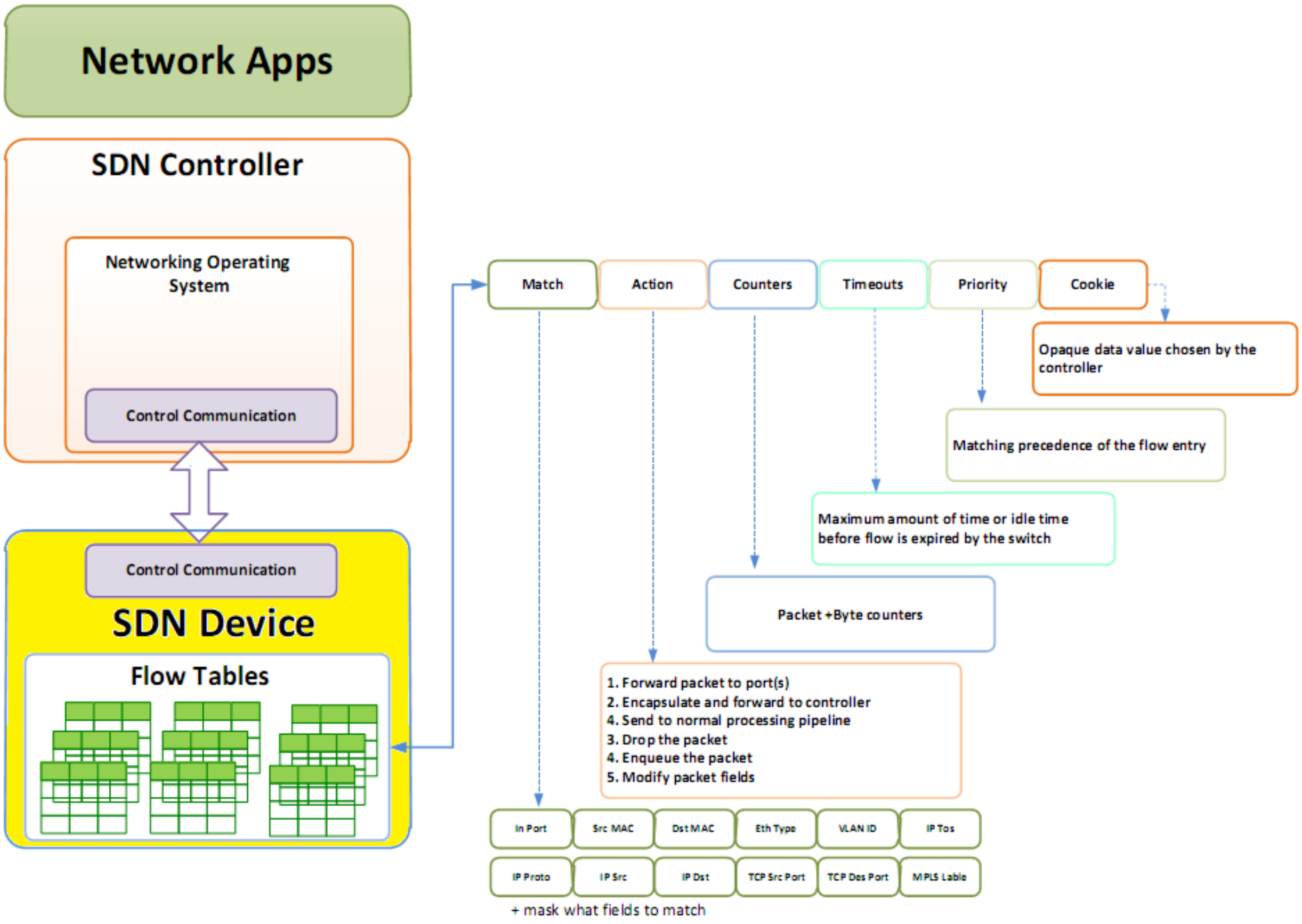

**Fig.2 : Architecture related to SDN considering OpenFlow**

The evolution of devices and mobile accessories, server virtualization and the emergence of cloud computing services has led to a re-review of the common architecture of networks[15]. Architecture of many traditional networks is hierarchy that is formed by using nodes of Ethernet switches in a tree structure[16]. This architecture will become clearer when the issue of client / server communications to be raised. But such a static architecture is not sufficient for dynamic communications and the needs of companies in the area of data centers and server media[17].

Computer networks covered the entire world and a serious and new development has not formed for many years. But companies and consumer organizations complain about the failure to add new features to their networks and according to their need, like much of the works to be done automatically and they also like to develop and expand their networks on the software side so they don't need new expensive hardware for many of the activities and needs of their network[18]–[20]. SDN architecture and Open Flow protocol make data and control levels separate from each other and networks smarter and more manageable and network infrastructure is isolated from applications. Companies will able to programming, network automation and more network control[9]. A new generation of SDN-based network using virtualization layers, the virtual switches, the central controller of communication standards and high-level API tries to do some of the tasks of monitoring and controlling switches and network routers in the upper layers by software[9]. In other words, SDN has reduced dependence on hardware and increases software capabilities and network intelligence. For years the network industry has not seen a new transformation and new standard has not introduced after TCP/IP protocol and transmission and control of data by switches and hardware routers. For three decades, hardware manufacturers are the undisputed rulers of this industry and every year introduce new models of their products with duplicate features "more speed" or "more security" into the market and small and large companies are also forced to buy and use these closed mysterious boxes[16], [19], [21].

## 1. Definition of the new architecture

When the network size increases, the number of requests to controllers in the SDN (software-based) network increases and the controller cannot answer all requests in a reasonable time. The delay time here is a delay in response to a data path request that has a significant impact on network latency. Also the purpose of reliability in SDN networks is that if the physical connection between the two nodes in the network is eliminated, the communication path between the nodes and their control is not eliminated. In this research, it is shown that the number of controllers and their location in SDN networks can affect the reliability and latency of SDN networks[22]. The relationship between controllers can be understood by adopting methods such as hierarchical methods. The topology in these networks is very noticeable and determines the amount of latency because it indicates how the edges are connected to each other and the weight of the edges determines the amount of delay in each connection. Most spatial services, the user of the division of the area also use a hierarchical structure. The formation of a hierarchical structure based on the division of the area into different regions and the selection of a set of these areas is done for the formation of areas with higher levels.

In this paper, the aim is to establish balance between reliability and latency [22] with proper placement of controllers in the SDN network in such a way that controllers' placement in these networks improves the reliability criterion without bringing unacceptable delay of the switch to controller.

## 2. Related works

In a [23], on the performance of the SDN controller using a number of policies available for Open Flow Controller is offered. The purpose of this paper is to provide a better understanding of the performance of the controller in SDN architecture. The purpose of this paper is to show that the work matter is not a fundamental limitation of the SDN network and in order to prove it, a multithreaded controller called NOX-MT has been suggested that is the new version of NOX controller and performance of NOX controller is improved thirty times. the goal is to answer this question that according to the network topology, how many controllers are needed and to determine the effect of the number of controllers on performance. With investigations, t is known that there is no specific law for each network, but the number of controllers depends on the network topology and selection of criteria.

In [24], Floodlight controller that is one of the Open Flow controllers has been examined and Controller affects the performance of the entire network. Basic concepts of SDN on Control Plan are done through a centralized controller. A centralized controller is a single entity that is responsible to manage all write devices.

In [7], a position controller algorithm and routing of traffic control in SDN networks has been proposed in order to maximize the ability to control traffic on SDN networks. In this architecture, a controller is used. This method cannot be held accountable for environments where more than one controller is required.

In [25], the reliability of traditional networks has been extensively investigated and models presented to them. The models presented in this research are not responsive to SDN networks.

In [18] three important issues have been focused in network. These issues include the existence of a mechanism that can be changed repeatedly according to network conditions and a high-level language exists to support network settings and a better view for monitoring network troubleshooting tasks as well as in diagnosis. For the solutions of these problems, there is a wide range of high-level language network policies.

In [26], A suitable framework for the development of multiple controller with WAN is suggested. This framework dynamically sets number of active controllers and total switches and ensures that the startup time and communication overhead are minimal and finally, optimization problem of supplying resources is formulized as a correct linear application.

## 3. Placement criteria

So far, we have introduced and compared latency definitions across the entire network and pointed to the importance of optimal positioning of controllers in order to reduce latency[22]. These issues are considered complex issues by considering k as the number of controllers for placement and also considering different value and importance consider different weight variations for the edges of the nodes (figure 3).

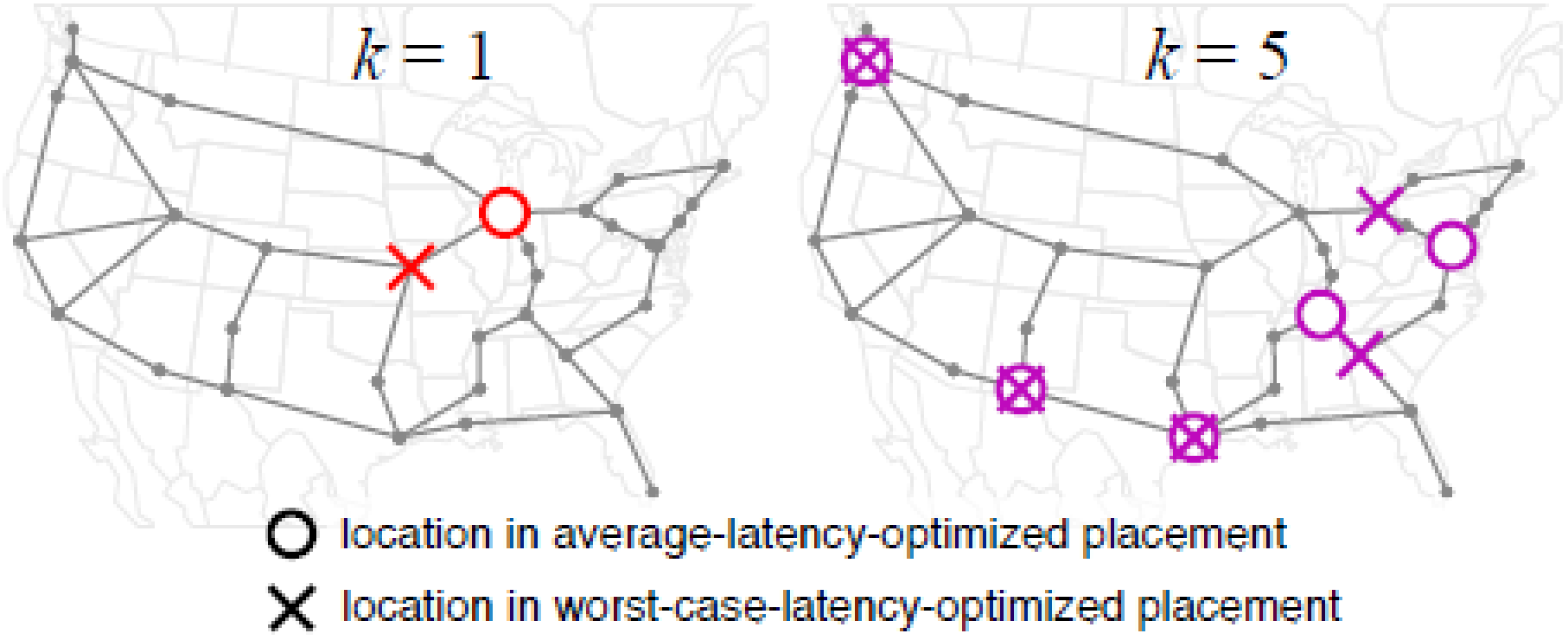

**Fig.3 : The network is tangibly represented by the graph G (V; E). When V is the set of nodes (switches) in the network and E is the set of edges between nodes (U; V) represents the physical link from node U to V. SDN control networks can be designed on network physics[22].**

It is also noteworthy that the control network such as the graph Gc (Vc; Ec) is considered in this regard that in this context Vc is a set of switches that are responsible for sending and receiving or distributing traffic control and Ec is a set of control paths. Hence, Ec includes control paths between switches and their controllers and is a complete mesh of control paths between all controllers. Considering a control grid and a specific routing mechanism for controlling traffic, it is possible to create a control path to the sequence of physical links and nodes in Puv and Pvu paths. In addition, $P_{\in c}$ can be used to specify a set of switches and links in the control path of $e_c$ When $e_c \in Ec$ .

- Average-case latency

Average-case latency for a network graph is represented by G (V; E) where the edge weight indicates the delay caused and (v;s)d represents the shortest path from the node $v \in V$ to $s \in V$ and number of nodes is n=|V| . Average-cage latency for positioning the controllers $S'$ is as follows:

In the optimization problem of the minimum k [22] the purpose of positioning $S'$ the set of all possible locations for the controller S such that K=| $S'$ | and Lavg=( $S'$ ) are minimum.

$$L_{avg}(S') = \frac{1}{n} \sum_{v \in V} \min_{(s \in S')} d(v,s) \qquad (1)$$

- Worst-case latency

Worst-case latency is one of the alternative criteria. In this method, the latency that is defined as the maximum node-to-controller is calculated as follows. In formula (2)

$$L_{\omega c}(S') = \max_{(v \in V)} \min_{(s \in S')} d(v,s) \qquad (2)$$

In this way, minimizing $S'$ is one of the desired goals and minimizing the optimization problem is related to the k number. Both methods; average-case latency and worst-case latency are calculated taking into account the distance from the node to the controller.

- Nodes within a latency bound

This method in contrast to two methods; average-case latency and worst-case latency seeks to locate the controllers in order to maximize the number of nodes at a delay time. One of these problems is the number k and the set of S = S1, S2, S3 ..., Sm Where $S'$ is subset of S = S1, S2, S3 ... Sm. In this approach, the goal is to identify and find the subset $S' \subseteq S$ of the set. In this case, U $S'$ is at most of K=| $S'$ | Each set of $Si$ contains all nodes with a delay bound for a single node. In most activities optimal positioning is done by direct measurements on all combinations of possible controllers. In order to analyze large networks, the problem of positioning is accomplished by the heavy weight of the solution between time and quality.

- **An outline of the proposed methodology**

Control networks for SDNs often have different shapes. In this regard, a lot of researches have been done for optimal placement and solving this problem. In most of these articles, control networks are classified as hierarchical networks, in other words by considering several controllers that communicate with each other in a mesh in the control network. Delay in the network can be determined by definite and probable criteria. Two crucial criteria can undoubtedly be the connection and coherence of the graph in the network which specifies respectively minimum cardinality between the node cut-set and edge cut-set. Nevertheless, definitive criteria usually do not work well. In general, the criteria for delays can be divided into two categories of connection criteria and traffic criteria. Based on connectivity, capabilities have focused on topology and network connectivity. One of the common connection criteria is k-terminal latency which is likely to be considered as a subset of k of nodes that can communicate with each other. In other words, traffic is a criterion for determining network performance. Traffic-based metrics are based on traffic capacity such as the number of service disruptions that result from failures can determine this quantity. For a SDN network the main function of the network is to control traffic distribution between control paths between different devices from the control level and data level. There is good communication between the various controllers on an SDN network without any defects in the valid paths between its switches and its controllers. This means that the switch can behave in the right direction. If the control path is disconnected this disconnect between the sending device and controllers or between different controllers will disrupt the network function. As a result, more control paths are faced with failure as well as further interruptions in traffic control. Therefore, traffic criteria are very understandable for introducing latency in SDN networks. Since no details of traffic control patterns are available, any control path that carries a unit of traffic and therefore uses a modified version of traffic criteria to describe the delay in the SDN control network. The delay criterion for most tasks performed is considered as the percentage of lost control paths; when the lost control path is identified by the control paths violated due to network failures.

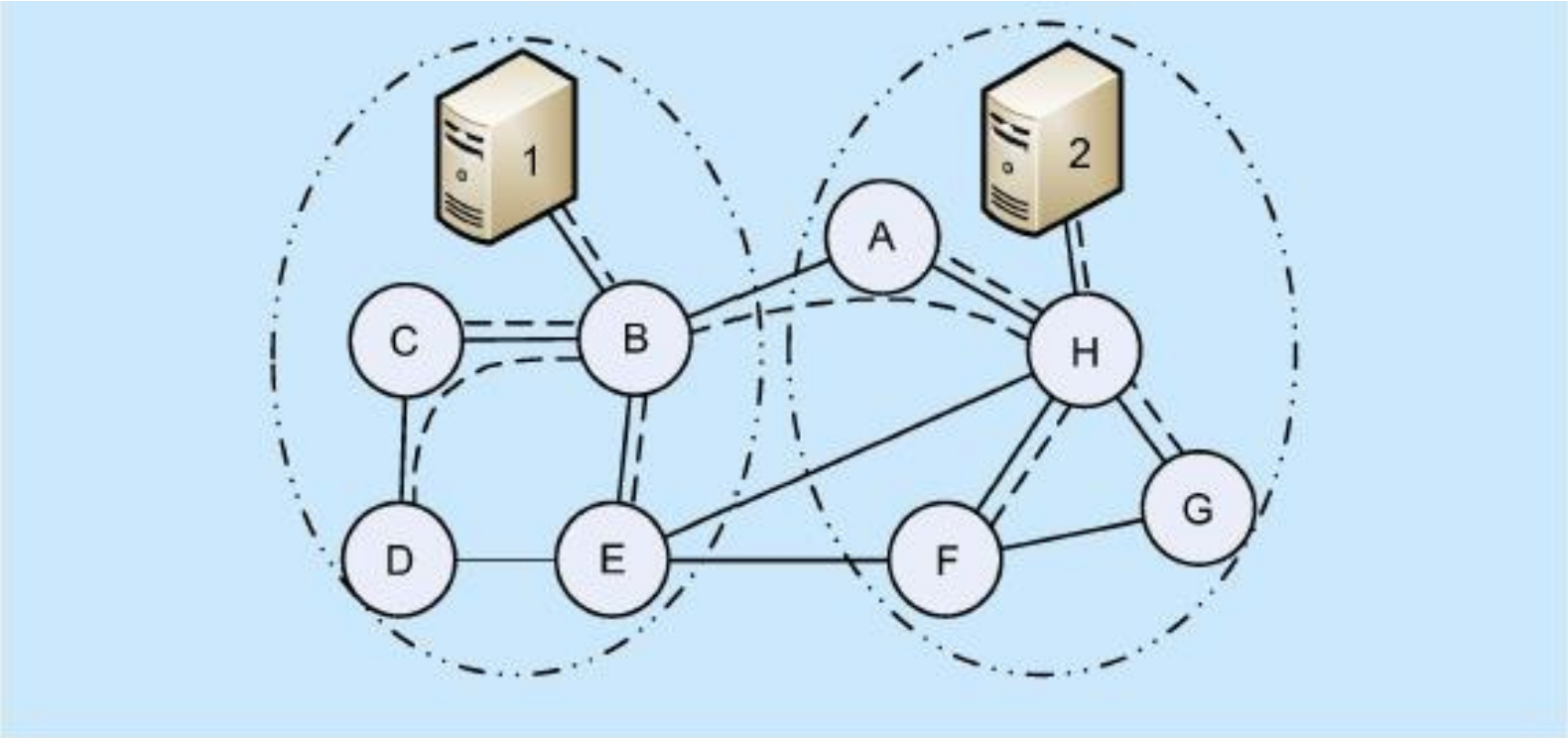

Fig.4 : An example of an SDN control network is divided into two categories

In figure 4 the bold lines represent the physical links and the dotted points represent the control paths. The goal of optimization is to minimize missing control paths. In SDN networks, delay analysis of a complex problem is due to the correlation between control paths. In a large number of cases two control paths may share a link or switch and as a result the dependence of the control paths will be unsuccessful. Most of these failures are related to the components of the control network and control paths which are due to physical causes and can be identified. In order to do this, first we will identify all the major failures in network physics. These failures happen independently. Each of these failures may affect many modes of control in the control network. Therefore, we consider s as a set of all network states that include failure states.

In this set $e_c \in E_c$ , $f_s$ is a set of disagreements caused by physical components. $f_s$ is a subset of VUE. In addition, $p_s$ shows the probability that the set s occurs which can be deduced from the previous operation data of the network and can easily show that:

$$\sum_{s\in S} p_s = 1 \tag{3}$$

When S occurs, the control paths between the switches and their controllers will fail. When $f_s$ separates the network, the control paths fail. For example, in figure 4 the failure of the H switch will separate controller 2 from the network. All of the control paths of the H switch to the controller 2 fail. Also due to the failure of the switch C the switch D separates the network which breaks the control path (B;D). In other words, if the network has not been scalable, the control paths affected can succeed or fail based on the failure mechanism used to control the traffic in the control of the track protection or routing. For example, in figure 4 if the IP link between the (A, H) does not succeed the (B, H) and (A, H) control paths will not succeed. The previous configuration of backlinks in the switches makes it possible to re-route the traffic control between a switch and its controller when there is an unsuccessful occurrence of the initial control paths. We can also model the effect of network failures on control paths as probability variables. Where is the formula 4:

$$q_s = \left[ control \ \ path \ \ e_c \ \ fails \middle| F_s \ \ in \ \ scenario \ \ s \, fails \right] \tag{4}$$

When $e_c \in E_c$ , $s \in S$ And $p_e \cap F_s = 0$ . $q_s$ is associated with the failure mechanism used in the control paths. In the remainder of the discussion, $q_s$ is assumed to be the same for all the controlled paths in the set s. In addition, in a set of failures all elements in the network physics G(V, E) are in definite and individual cases. Each switch identifies the failures and routing of traffic re-control independently. Therefore, the failures are conditional on independent control paths and latency analysis techniques are used with independent failures. For each set of failures a number of unsuccessful control paths and delay criteria and lost acceptable control paths can be obtained. In the following we reach the graph which leads to the formulation of the problem. The algorithm we propose is to find nodes that are commensurate with the cost function of the problem (this research emphasizes the delay) which has the lowest average cost. In addition to this algorithm, one of the general or local optimization algorithms can be used. If the dimensions of the graph are small then general optimization methods are used and if the dimensions of the graph are large we will use local optimization methods. For the number of controllers the next parameter is important that by partitioning the components of the graph, the number of controllers in each section will also be determined. Finally, placement results of the random number will be compared with the general and local optimal state.

## 4. Evaluation of proposed method

A simple network design was done manually in gephi software with 23 hosts and lots of connections. The following figure shows this network. As shown in figure 5, four groups of hosts are considered. These four groups can be considered as four distinct cities that all of which together form an interconnected network. Of course when this representation is performed by a computer, this fact may be less visible. The purpose of the manual drawing of this network is also to better understand the idea of this research. Drawing of this network in MATLAB software is shown in figure 6.

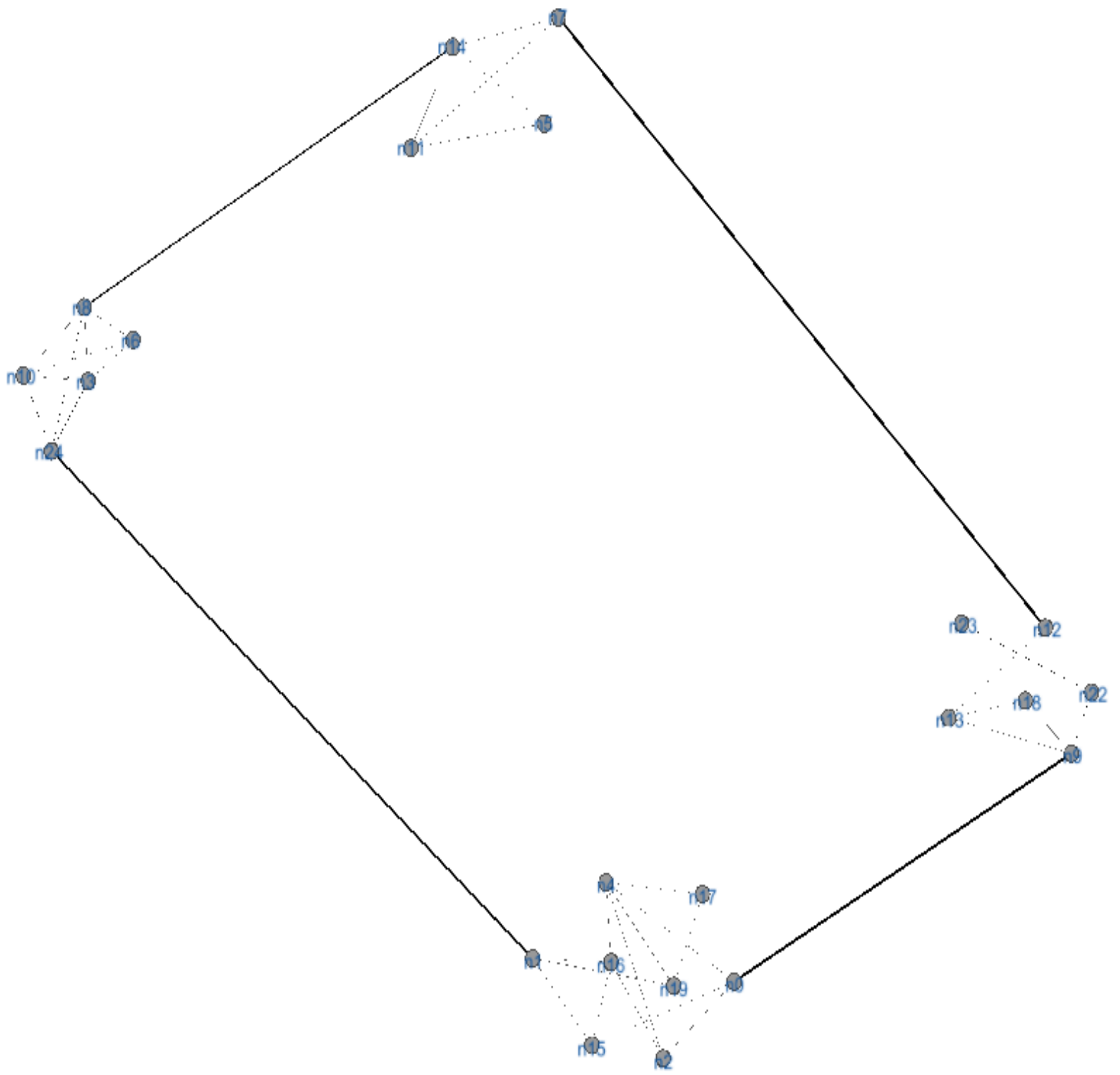

**Fig.5 : Analyzed network in this research**

As stated above, different categories are included in the design of the network. Connections between different hosts have different delays. When plotting the network, these delays are displayed in terms of connection weight (cost). Gephi software when designing a network allows displaying the diameter of the edges to fit into the weights of connections. Therefore, thin lines represent low-latency connections and large lines represent high latency connections. The cost of connections in each category is less than the connections between different categories. For calculations, delays were considered unilaterally and between one to one hundreds which would not create a gap in the whole problem. In Figure 7 four groups are shown separately.

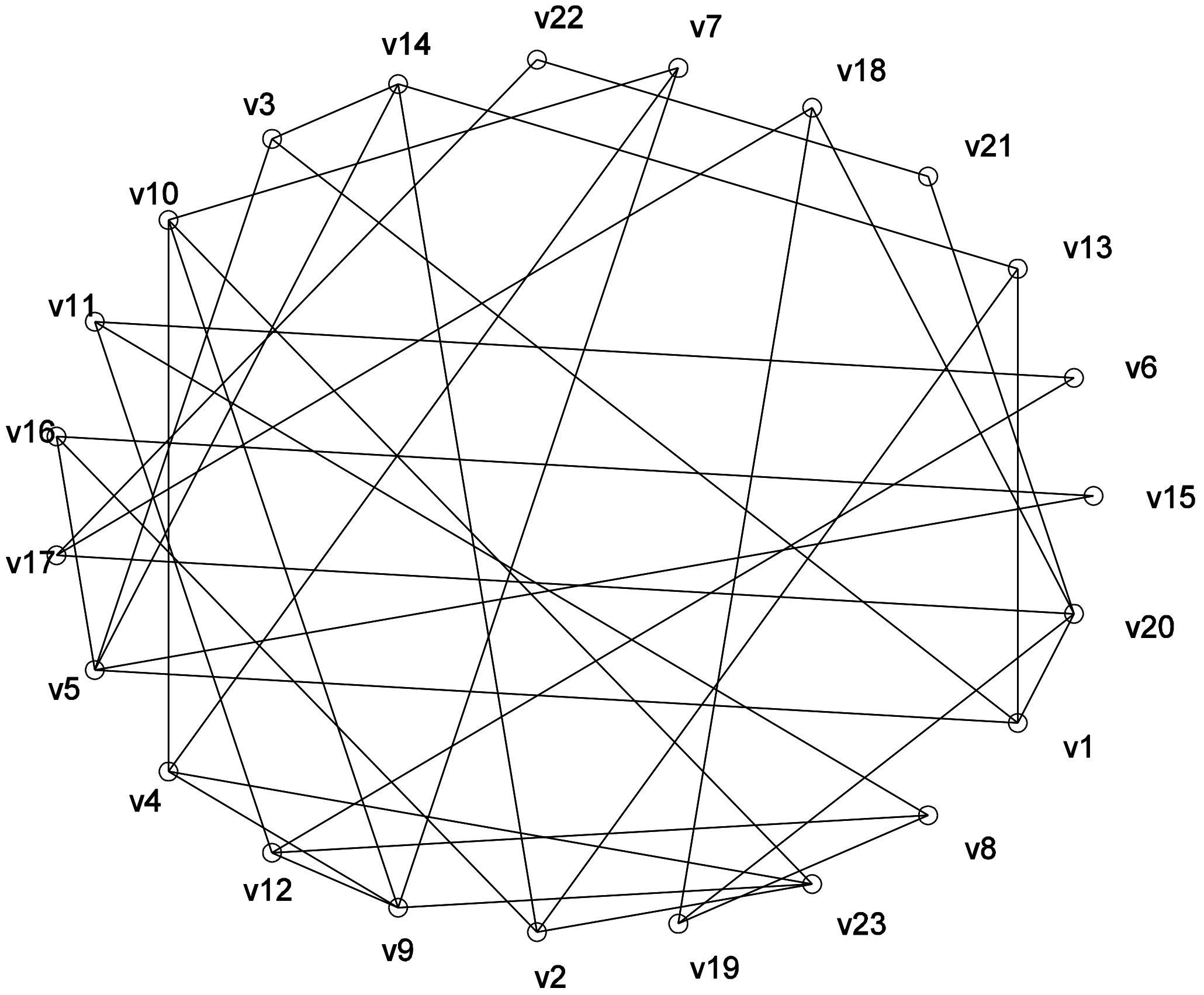

**Fig.6 : Draw (Analyzed network figure 5) in MATLAB software.**

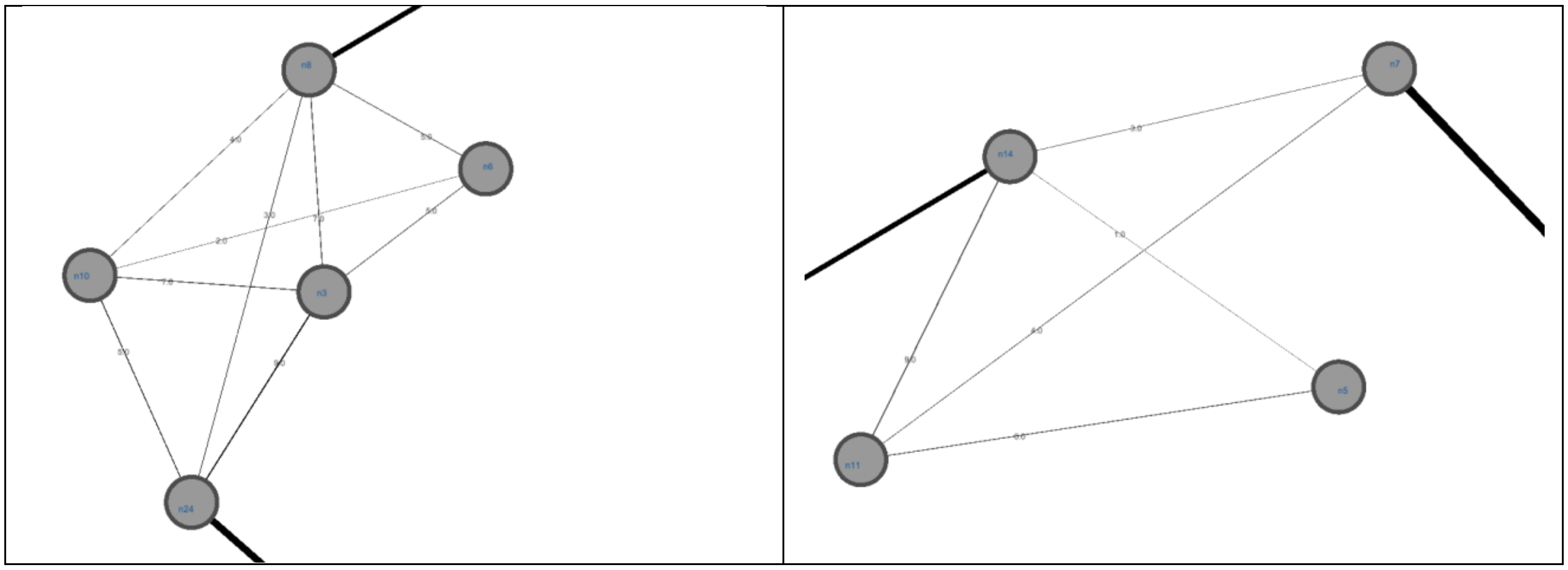

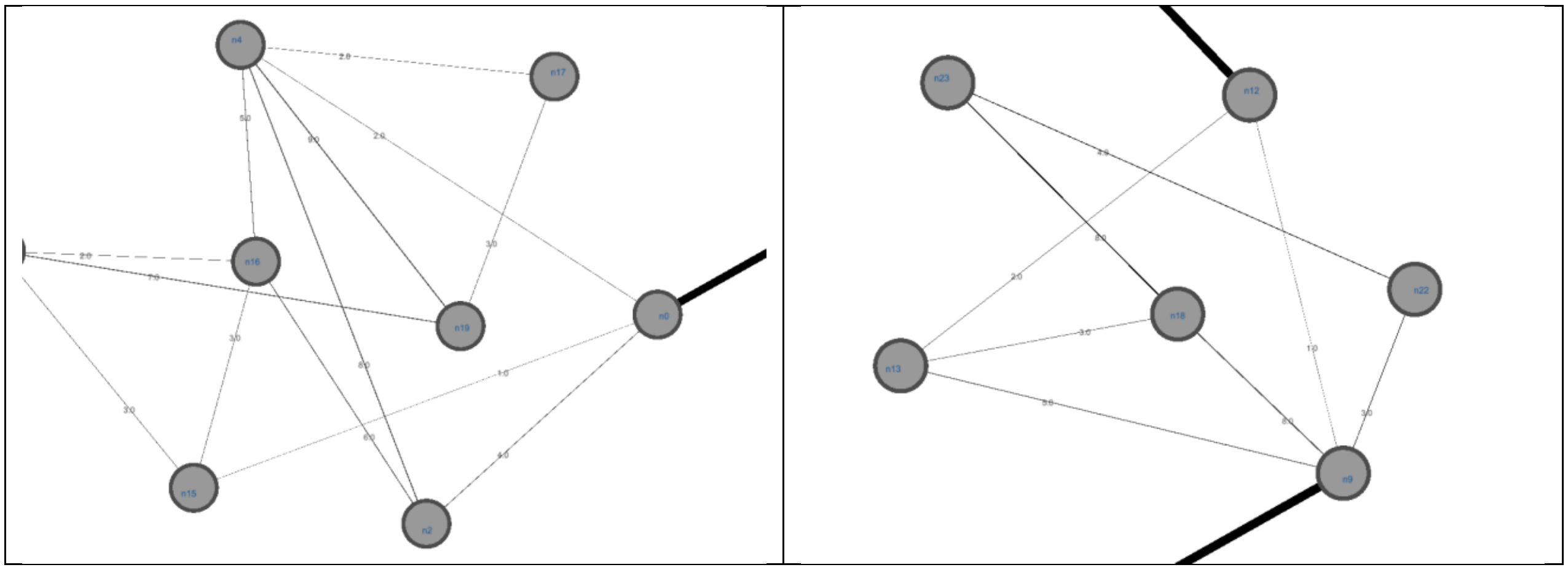

**Fig.7 : Connections between different hosts**

After designing the network in Gephi as desired, it should be prepared in a way for subsequent processing in MATLAB. The most common way to represent undirected weighted networks is to use adjacency matrix. The un-directional purpose of the network is that when a host can contact another host then the second host can contact the first host that indicates un-directional network. The Gephi output is first stored as a file “comma separated values” and then will be downloadable from MATLAB. After loading the CSV file in MATLAB, a matrix is obtained that its rows and columns represent the hosts and its layers represent the latency of the connections.

### 5. Optimization criteria including delay and cost

- **First view: A controller in the ideal place (k = 1)**

The first view was to significantly reduce the number of controllers and only use a controller in the network. This view looks at the issue only economically and hardwarely. In this way, the optimal location is used to answer the second question. It means the delay of all hosts to each other and the hosting that has the least latency of all other hosts is chosen as the optimal location for the controller. Table 1 describes the optimal host as well as the total delay to nodes for the designed network in the previous step.

Table 1: Host for placement of controller

| Suitable host for controller placement | Total delay to all hosts |
|---|---|
| **Host** | 1290 |

Since the number of controllers in this case is only one the average delay between all controllers does not mean a specific meaning. This explanation has since been mentioned that in the next scenarios between different controllers, the average latency of access to subsequent or following hosts will be calculated.

- **Second view: A certain number of controllers in different locations**

In this case, before the discussion about the location of controllers, their number is already specified. In this case these controllers are placed on arbitrary nodes and then the delay caused by the connection of these controllers with the subsequent hosts is calculated. In order to determine the subsequent or following hosts of a particular controller, the simple approach of the nearest neighbor is used. That is a host follows a controller if it has the lowest latency for all other controllers. Figure 8 illustrates the details of this method

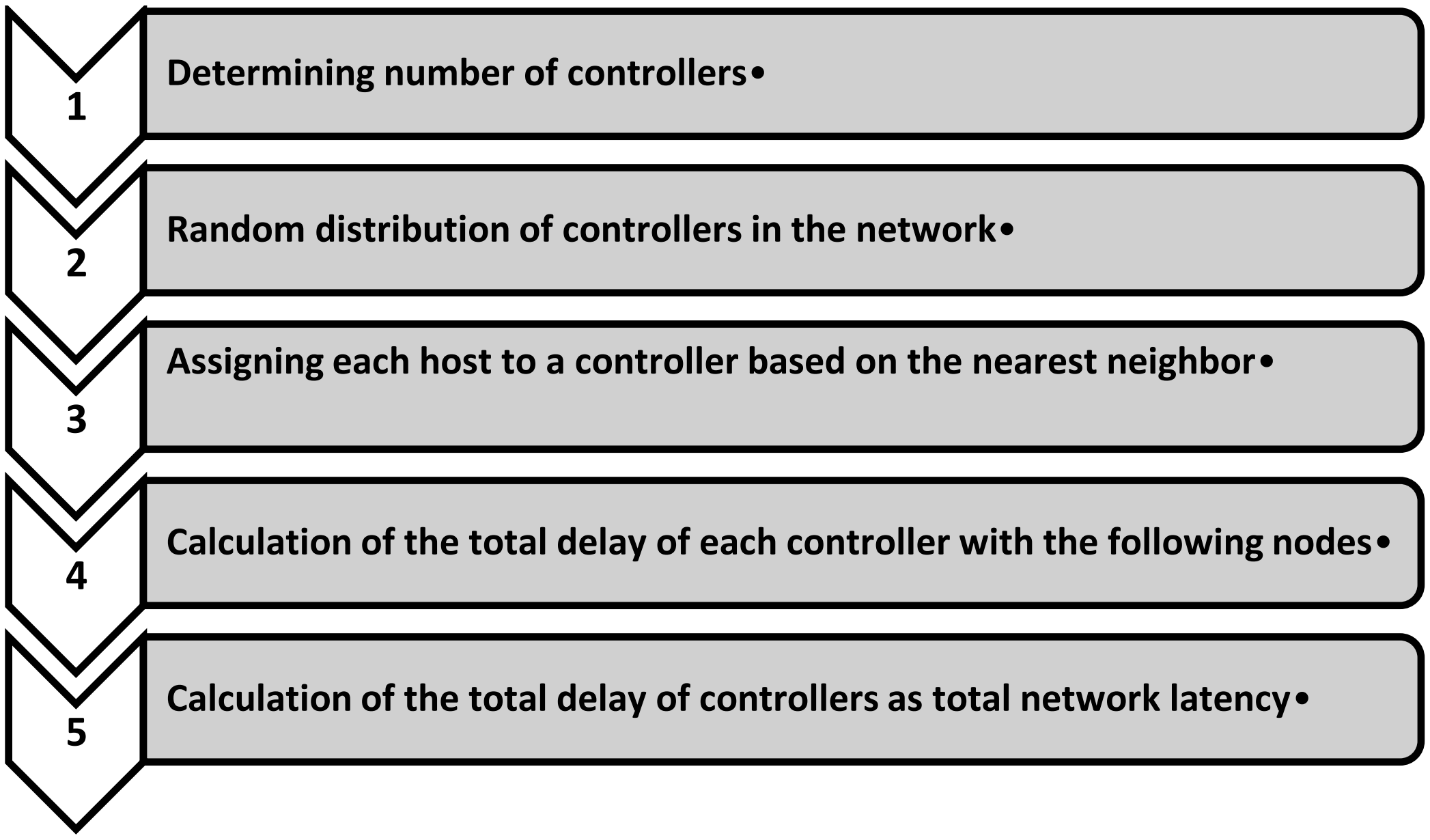

**Fig.8 : Flow chart of the second view.**

This method was tested for one to ten controllers in the designed network. Since the distribution of controllers in the network is random, in different performances the answers will not be completely identical. Of course, if the number of hosts is much larger than the number of controllers (this is not the case in the example network) then the results will be the same. Table 2 shows the numerical value of Fig. 9.

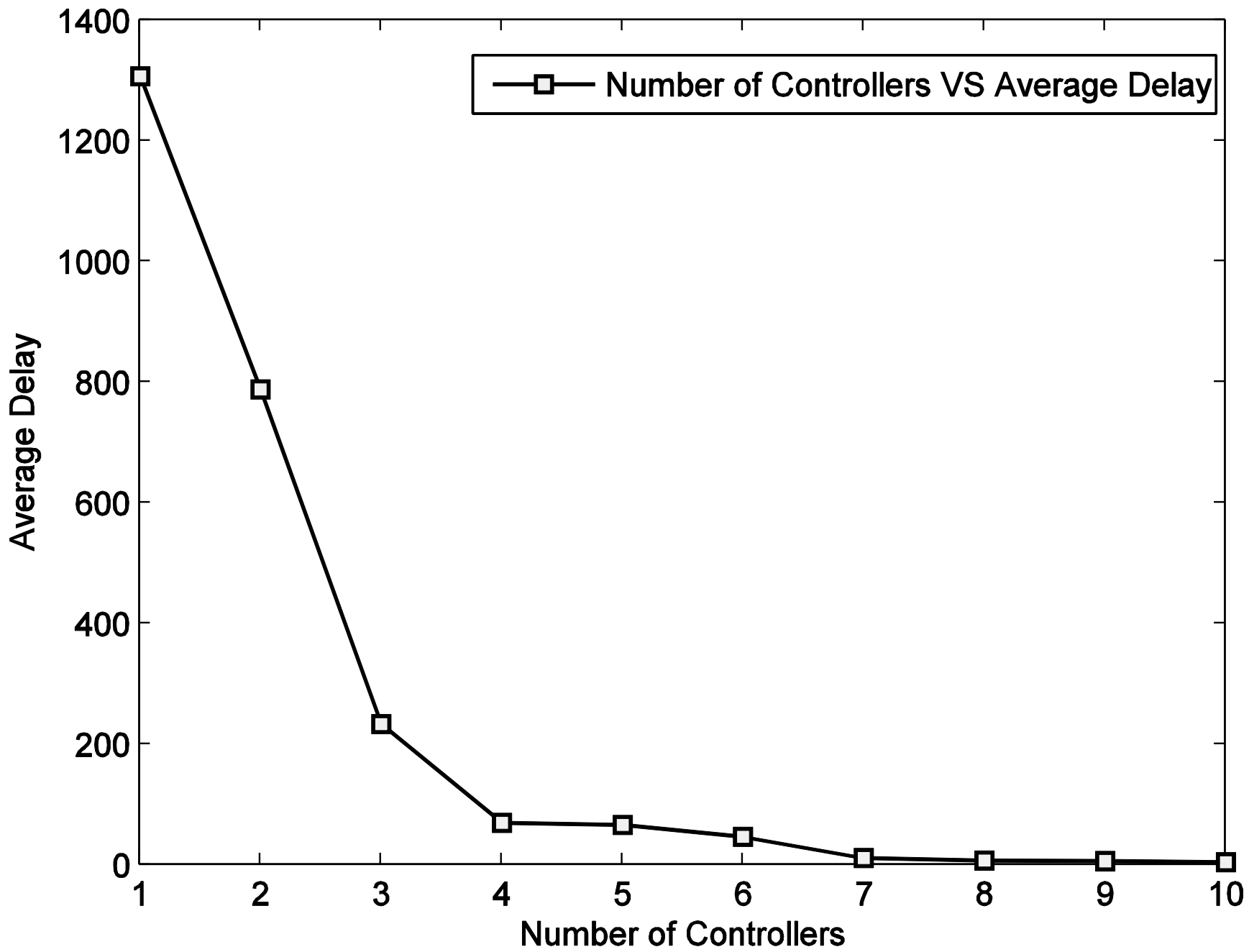

Fig 9 : The average delay compared to the change in the number of controllers

Table 2: The numerical value of Fig. 9

| Number of controllers | delay average | decrease of delay average |
|---|---|---|
| **1** | 1291 | - |
| **2** | 389 | 902 |
| **3** | 302 | 87 |
| **4** | 188 | 114 |
| **5** | 134 | 54 |
| **6** | 48 | 86 |
| **7** | 12 | 36 |
| **8** | 7 | 5 |
| **9** | 5 | 2 |
| **10** | 5 | 0 |

What can be deduced from the numbers in Table 2 is that with the increase in the number of controllers, the delay average sharply decreased but if this process increases, the deceleration of the delay average will also decrease sharply. This fall in the average delay rate clearly states that for each network there is probably an optimal controller number which if placed in suitable situations, the increase in the number of controllers will not significantly reduce the average of delay.

- **A third view: Using the optimal number of controllers in optimal locations**

The weakness of the two previous perspectives which is usually the most commonly used approach to current software networks is easily apparent or visible in facing delay. In the first place, despite the cost reduction and simplicity in designing network management algorithms, the delay is considerably high so that it is not acceptable for many networks with a large number of hosts. Considering the increased cost of having multiple software controllers, second method reduces the delay but does not consider the location factor as well as its number. For this reason, there is no balance between the number of controllers and the network latency. For example, in the delay table with the increase in the number of controllers, after increasing a certain number of controllers, the reduction of the delay is not significant and adding to the cost of the controllers does not have the benefit of reducing the delay. In addition, there has been no discussion of the location of controllers. Therefore, the delays with this number of controllers are not optimal and can be reduced by moving the controller's location. As shown in Figure10 the increased latency is shown due to the lack of proper location. After implementing the proposed algorithm on the network, the number of controllers, four numbers and their location is shown in the table 3.

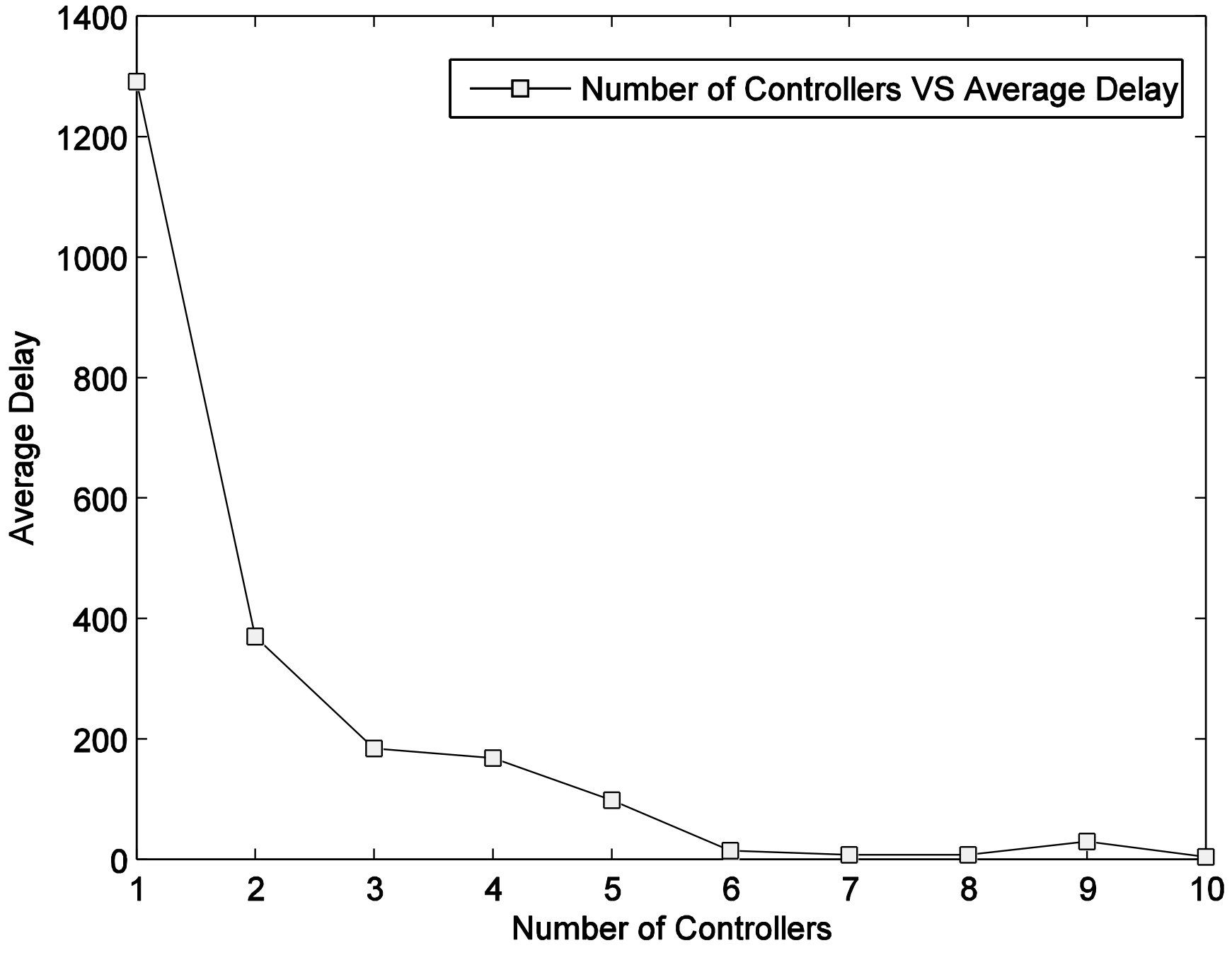

**Fig 10 : Delays in the number of controllers**

Table 3: The number of optimal controllers, four numbers and their location

| Controller location (managing host) | Following nodes | Total delay to neighbors in the cluster |
|---|---|---|
| **1** | 0,2,4,12,13,14,15 | 26 |
| **10** | 5,7,11 | 18 |
| **6** | 3,8,9,22 | 11 |
| **19** | 16,17,18,20,21 | 20 |

Finally, the average delays with the presence of four controllers were equal to 18/75which is much less than the latency achieved with the implementation of previous methods. In addition, the optimal number of controllers is also used.

## 6. Conclusions and future works

In this research, a method was presented for determining the number and location of controllers in software networks. There were two common practices in the previous implementation and simulation. In general only one controller is used and most of the proposed algorithms are designed to control the network traffic conditions according to the assumption of a controller that is aware of all network conditions. The great disadvantage of this view is the high latency of communication. In other methods, there was no broad discussion of the number and location of optimal controllers in the network. But in this study, these two issues were deeply studied and simple but effective ideas were presented. The backbone of the proposed method is network clustering. Various clustering methods are presented in computer networks. Since the focus of the current research was on reducing delays, Clustering was designed to fit the delay of connections. The first step in extending and generalizing the proposed method is to add many other factors that affect the network. An important factor that is often discussed in the Internet network and forms the basis of its construction is reliability. This parameter discusses whether with the loss of some connections, the proper management function of controllers in the network can still be trusted.